# Helicity-dependent optical control of the magnetization state emerging from the Landau-Lifshitz-Gilbert equation

Benjamin Assouline, Amir Capua*

Department of Applied Physics, The Hebrew University of Jerusalem, Jerusalem 9190401, Israel

*e-mail: amir.capua@mail.huji.ac.il

**Abstract:**

It is well known that the Gilbert relaxation time of a magnetic moment scales inversely with the magnitude of the externally applied field, $H$, and the Gilbert damping, $\alpha$. Therefore, in ultrashort optical pulses, where $H$ can temporarily be extremely large, the Gilbert relaxation time can momentarily be extremely short, reaching even picosecond timescales. Here we show that for typical ultrashort pulses, the optical control of the magnetization emerges by merely considering the optical magnetic field in the Landau-Lifshitz-Gilbert (LLG) equation. Surprisingly, when circularly polarized optical pulses are introduced to the LLG equation, an optically induced helicity-dependent torque results. We find that the strength of the interaction is determined by $\eta = \alpha\gamma H/f_{opt}$, where $f_{opt}$ and $\gamma$ are the optical frequency and gyromagnetic ratio. Our results illustrate the generality of the LLG equation to the optical limit and the pivotal role of the Gilbert damping in the general interaction between optical magnetic fields and spins in solids.

The ability to control the magnetization order parameter using ultrashort circularly polarized (CP) optical pulses has attracted a great deal of attention since the early experiments of the all-optical helicity dependent switching (AO-HDS) [1-4]. This interaction was found intriguing since it appears to have all the necessary ingredients to be explained by a coherent transfer of angular momentum, yet it occurs at photon energies of $1 - 2 \, eV$, very far from the typical resonant transitions in metals. The technological applications and fundamental scientific aspects steered much debate and discussion [5,6], and the experiments that followed found dependencies on a variety of parameters including material composition [7-9], magnetic structure [10-12], and laser parameters [1,3,13], that were often experiment-specific [4]. Consequently, a multitude of mechanisms that entangle photons [14,15], spins [16,17], and phonons [18,19] have been discovered. References [4,20] provide a state of the art review of the theoretical and experimental works of the field.

Ferromagnetic resonance (FMR) experiments are usually carried out at the $GHz$ range. In contrast, optical fields oscillate much faster, at $\sim 400 - 800 \, THz$. Therefore, it seems unlikely that such fast-oscillating fields may interact with magnetic moments. However, the amplitude of the magnetic field in ultrashort optical pulses can, temporarily, be very large such that the magnetization may respond extremely fast. For example, in typical experiments having $40 \, fs - 1 \, ps$ pulses at $800 \, nm$, with energy of $0.5 \, mJ$ that are focused to a spot size of $\sim 0.5 \, mm^2$, the peak magnetic flux density can be as high as $\sim 5 \, T$, for which the corresponding Gilbert relaxation time reduces to tens of picoseconds in typical ferromagnets.

Here we show that ultrashort optical pulses may control the magnetization state by merely considering the optical magnetic field in the Landau-Lifshitz-Gilbert (LLG) equation. We find that the strength of the interaction is determined by $\eta = \alpha \gamma H / f_{opt}$, where $f_{opt}$ and $\alpha$ are the angular optical frequency and the Gilbert damping, respectively, and $\gamma$ is the gyromagnetic ratio. Moreover, we show that for circularly polarized (CP) pulses, the polarity of the optically induced torque is determined by the optical helicity. From a quantitative analysis, we find that a sizable effective out-of-plane field is generated which is comparable to that measured experimentally in ferromagnet/heavy-metal (FM/HM) material systems.

The LLG equation is typically not applied in the optical limit, and hence requires an alternative mathematical framework whose principles we adopt from the Bloch equations for semiconductor lasers [21,22]. We exploit the analogy between the magnetization state and the Bloch vector of a two-level system (TLS) [23,24] by transforming the LLG equation under a time-varying magnetic field excitation to the dynamical Maxwell-Bloch (MB) equations in the presence of an electrical carrier injection. In this transformation, the reversal of the magnetization is described in terms of population transfer between the states.

The paper is organized as follows: We begin by transforming the LLG equation to the density matrix equations of a TLS. We then identify the mathematical form of a time-dependent magnetic field in the LLG equation, $\vec{H}_{pump\downarrow\uparrow}$, that is mapped to a time-independent carrier injection rate into the TLS. Such excitation induces a population transfer that varies linearly in time and accordingly to a magnetization switching profile that is also linear in time. The mathematical $\vec{H}_{pump\downarrow\uparrow}$ field emerges naturally as a temporal impulse-like excitation. We then show that when $\alpha$ is sizable, $\vec{H}_{pump\downarrow\uparrow}$ acquires a CP component whose handedness is determined by the direction of the switching. By substituting $\vec{H}_{pump\downarrow\uparrow}$ for an experimentally realistic picosecond CP Gaussian optical magnetic pulse, we show that it can also exert a net torque on the magnetization. In this case as well, the helicity determines the polarity of the torque. Finally, we present a quantitative analysis that is based on experimental data.

The LLG equation describing the dynamics of the magnetization, $\vec{M}$, where the losses are introduced in the Landau–Lifshitz form is given by [25]:

$$\frac{d\vec{M}}{dt} = -\frac{\gamma}{1+\alpha^2}\vec{M}\times\vec{H} - \frac{\gamma\alpha}{1+\alpha^2}\frac{1}{M_s}\vec{M}\times\vec{M}\times\vec{H}. \qquad (1)$$

Here $M_s$ and $\vec{H}$ are the magnetization saturation and the time dependent externally applied magnetic field, respectively. We define $\vec{H}_{eff}$ by:

$$\vec{H}_{eff} \triangleq \left(\vec{H} - \frac{\alpha}{M_s}\vec{H}\times\vec{M}\right), \qquad (2)$$

and in addition, $\kappa \triangleq \frac{\gamma}{1+\alpha^2}(H_{eff\,x} - jH_{eff\,y})/2$ and $\kappa_0 \triangleq \frac{\gamma}{1+\alpha^2}H_{eff\,z}$, where $\kappa$ and $\kappa_0$ can be regarded as effective AC and DC magnetic fields acting on $\vec{M}$, respectively. We transform $\vec{M}$ to the density matrix elements of the Bloch state in the TLS picture and compare it to the Bloch equations describing a semiconductor laser that is electrically pumped [26]:

$$\begin{cases} \dot{\rho}_{11} = \Lambda_1 - \gamma_1 \rho_{11} + \frac{j}{2}[(\rho_{12} - \rho_{21})(V_{12} + V_{21}) - (\rho_{12} + \rho_{21})(V_{12} - V_{21})] \\ \dot{\rho}_{22} = \Lambda_2 - \gamma_2 \rho_{22} - \frac{j}{2}[(\rho_{12} - \rho_{21})(V_{12} + V_{21}) - (\rho_{12} + \rho_{21})(V_{12} - V_{21})] \\ \dot{\rho}_{12} = -(j\omega_{TLS} + \gamma_{inh})\rho_{12} + j(\rho_{11} - \rho_{22})V_{12} \end{cases} \quad (3)$$

In this reference model, $\Lambda_1$ and $\Lambda_2$ are injection rates of carriers to the ground and excited states of the TLS, respectively. They are assumed to be time independent and represent a constant injection of carriers from an undepleted reservoir [27]. $\gamma_1$ and $\gamma_2$ are the relaxation rates of the ground and excited states, and $\gamma_{inh}$ is the decoherence rate due to an inhomogeneous broadening. $V_{12}$ is the interaction term and $\omega_{TLS}$ is the resonance frequency of the TLS. Figure 1(a) illustrates schematically the analogy between the magnetization dynamics and the electrically pumped TLS. We find the connection between the LLG equation expressed in the density matrix form and the model of the electrically pumped TLS:

$$\begin{cases} \Lambda_1 - \gamma_1 \rho_{11} + [M_y \Re\{V_{12}\} + M_x \Im\{V_{12}\}] = -j\kappa\rho_{21} + c.c. \\ \Lambda_2 - \gamma_2 \rho_{22} - [M_y \Re\{V_{12}\} + M_x \Im\{V_{12}\}] = j\kappa\rho_{21} + c.c. \\ -(j\omega_{TLS} + \gamma_{inh})\rho_{12} + jM_z V_{12} = -j\kappa_0 \rho_{12} + j\kappa M_z \end{cases} \quad (4)$$

The pumping of the excited and ground states by the constant $\Lambda_1$ and $\Lambda_2$ rates implies that the reversal of the magnetization along the $\mp \hat{z}$ direction is linear in time. Using Eq. (4) we find $\kappa$, and hence a field $\vec{H}$, that produces such $\Lambda_1$ and $\Lambda_2$. We define this field as $\vec{H}_{pump\downarrow\uparrow}$:

$$\vec{H}_{pump\downarrow\uparrow} = \frac{\pm \Lambda_p}{M_s^2 - M_z^2} \begin{pmatrix} M_y \\ -M_x \\ 0 \end{pmatrix}. \quad (5)$$

$\vec{H}_{pump\downarrow\uparrow}$ depends on the temporal state of $\vec{M}$ while $\Lambda_p = \gamma \Lambda_1/(1 + \alpha^2)$ is the effective field strength parameter. $\vec{H}_{pump\downarrow}$ and $\vec{H}_{pump\uparrow}$ induce a linear transition of $\vec{M}$ towards the $-\hat{z}$ and $+\hat{z}$ direction, respectively.

Figure 1(b) presents the outcome of the application of $\vec{H}_{pump\downarrow\uparrow}$ by numerically integrating the LLG equation. The Figure illustrates $\vec{H}(t)$, $M_z(t)$, and the $\hat{z}$ torque, $\left(-\vec{M}\times\vec{H}\right)_z$, for alternating $\vec{H}_{pump\downarrow}$ and $\vec{H}_{pump\uparrow}$ that switch $\vec{M}$ between $\mp M_s\hat{z}$. The magnitude of $\Lambda_p$ determines the switching time, $\Delta\tau_{\downarrow\uparrow}$, chosen here to describe a femtosecond regime. Equation (4) yields $\Delta\tau_{\downarrow\uparrow} = (1+\alpha^2)M_s/(\gamma\Lambda_p) \approx M_s/\gamma\Lambda_p$ in which $M_z$ is driven from $M_z = 0$ to $M_z \cong \pm M_s$ (for derivation, see Supplemental Material Note 1). It is seen that $\left(-\vec{M}\times\vec{H}\right)_z$ is constant when $\vec{H}_{pump\downarrow}$ or $\vec{H}_{pump\uparrow}$ are applied so that the switching profile of $M_z$ is linear in time. It is also seen that $\vec{H}_{pump\downarrow\uparrow}$ requires that $|\vec{H}|$ diverge as $M_z$ approaches $\pm M_s$, which is not experimentally feasible. To account for a more realistic excitation, in Fig. 1(c) we simulated a pulse whose trailing edge was taken as a reflection in time of $\vec{H}_{pump\downarrow\uparrow}$, and that is shorter by an order of magnitude as compared to the leading edge. In this case $\vec{M}$ remains in its final state when $\vec{H}$ is eventually turned off.

The polarization state of $\vec{H}_{pump\downarrow\uparrow}$ is determined from the polarization state of the transverse components of $\vec{M}$. Next, we show that for larger $\alpha$, $M_y(t)$ becomes appreciable such that $\vec{H}_{pump\downarrow\uparrow}$ acquires an additional CP component. This result emerges naturally from the Bloch picture: we recall that the transverse components of $\vec{M}$ are expressed by the off-diagonal density matrix element. According to Eq. (3), $\rho_{12}$ oscillates at $\omega_{TLS}$ and decays at the rate $\gamma_{inh}$, whereas the sign of $\omega_{TLS}$ determines the handedness of the transverse components of $\vec{M}$. Namely, the ratio between $\omega_{TLS}$ and $\gamma_{inh}$ determines the magnitude of the circular component in the $\left(M_x(t), M_y(t)\right)$ trajectory. Under the application of $\vec{H}_{pump\downarrow\uparrow}$, Eq. (4) yields $\omega_{TLS} = \pm\gamma\Lambda_p\alpha M_s/[(M_s^2 - M_z^2)(1+\alpha^2)]$ and $\gamma_{inh} = \mp\gamma\Lambda_p M_z/[(M_s^2 - M_z^2)(1+\alpha^2)]$ readily showing that $|\omega_{TLS}/\gamma_{inh}| = \alpha M_s/M_z$ increases with $\alpha$, so that $\vec{H}_{pump\downarrow\uparrow}$ acquires an additional CP component (see Supplemental Note 2 for full derivation). Figure 2 illustrates these results. Panel (a) presents the components of $\vec{M}(t)$ for the same simulation in Fig. 1(b). It is seen that $M_y(t)$ is negligible and thus $\vec{H}_{pump\downarrow\uparrow}$ remains linearly polarized. When $\alpha$ is increased, an elliptical trajectory of $\vec{M}$ in the $x-y$ plane emerges, while the constant transition rate of $M_z$ persists as illustrated in Fig. 2(b). In this case, $\vec{H}_{pump\downarrow\uparrow}$ acquires a right-CP (RCP) or left-CP (LCP) component depending on the choice of $\vec{H}_{pump\downarrow}$ or $\vec{H}_{pump\uparrow}$.

The coupling between the handedness and reversal direction in a femtosecond excitation is reminiscent of the switching reported in AO-HDS experiments and emerges naturally in our model. These results call to examine the interaction of the CP magnetic field of a short optical pulse with $\vec{M}$. Figure 3(a) presents the calculation for experimental conditions [4]. The results are shown for an $800 \, nm$ optical magnetic field of an RCP Gaussian optical pulse $\vec{H}_{opt}(t)$. The pulse has a duration determined by $\tau_p$, an angular frequency $\omega_{opt}$, and a peak amplitude $H_{peak}$ that is reached at $t = t_{peak}$. In our simulations $\tau_p = 3 \, ps$ and $t_{peak} = 10 \, ps$. The pulse energy was $\sim 5 \, mJ$ and assumed to be focused to a spot size of $\sim 100 \, \mu m^2$, for which $H_{peak} = 8 \cdot 10^6 \, A/m$. Here we take $\alpha = 0.035$ [28,29]. For such conditions, the Gilbert relaxation time corresponding to $H_{peak}$ is $\tau_\alpha = \frac{1}{\alpha \gamma H_{peak}} \approx 16 \, ps$ [30]. It is readily seen that for such $\tau_\alpha$ the magnetization responds within the duration of the optical pulse indicating that the interaction between the optical pulse and $\vec{M}$ becomes possible by the LLG equation. Following the interaction, $M_z = -5 \times 10^{-4} \cdot M_S$, namely a sizable net longitudinal torque results. In agreement with the prediction of the TLS model, pulses of the opposite helicity induce an opposite transition as shown in Fig. 3(b). The results are compared to the measured data discussed in Supplemental Material Note 3. To this end we simulate the same conditions of the measurements including optical intensity and sample parameters. Accordingly, we find from our calculations an effective field which is of the same order of magnitude as measured.

For a given pulse duration, we define the interaction strength parameter $\eta = 2\pi\alpha\gamma H_{peak}/\omega_{opt}$, which expresses the ratio between $\tau_\alpha$ and the optical cycle and is $2.5 \cdot 10^{-4}$ in Fig. 3(a). The principles of the interaction can be better understood at the limit where $\eta \to 1$ and for which the interaction can be described analytically. To this end, we set $\eta = 1$. The higher optical magnetic fields required for this limit are achievable using conventional amplified femtosecond lasers, for example by focusing a $\sim 5 \, mJ$ pulse into a spot size of $\sim 1 \, \mu m^2$. Figure 3(c) illustrates the results for an RCP $\vec{H}_{opt}$ pulse of a duration of $20 \, fs$ determined by the full width at half-maximum of the intensity. The Figure reveals the different stages of the interaction. During the leading edge, for $t < \sim 40\ fs$, the relative phase between $\vec{H}_{opt}$ and $\vec{M}$ seems arbitrary. As $t_{peak}$ is reached, the Gilbert relaxation time becomes as short as the optical cycle allowing $\vec{M}$ to follow $\vec{H}_{opt}$ until it is entirely locked to $\vec{H}_{opt}$. In this case, $\vec{M}$ undergoes a right-circular trajectory about $\hat{z}$. The switching of $\vec{M}$ takes place at the final stage of the interaction: During the trailing edge of the pulse, the amplitude of $\vec{H}_{opt}$ reduces and $\tau_\alpha$ extends, thereby releasing the locking between $\vec{M}$ and $\vec{H}_{opt}$. In this case, the switching profile of $M_z$ is monotonic linear-like in time, closely resembling the transition stemming from a constant carrier injection rate in the Bloch picture. The optically induced transition can be described analytically following the calculation presented in Supplemental Note 4, from which we find the transition rate:

$$\Gamma/M_s = \mp \frac{3}{2\sqrt{2}} \ln\left(\frac{4}{3}\right) \frac{1}{\tau_p \left[\sqrt{\ln\left(\frac{H_{peak}}{0.27 H_{th}}\right)} - \sqrt{\ln\left(\frac{H_{peak}}{H_{th}/\sqrt{2}}\right)}\right]}, \qquad (6)$$

where $H_{th} = \frac{\omega_{opt}}{2\pi\gamma\alpha}$ is the value of $H_{peak}$ at $\eta = 1$. The rate $\Gamma/M_s$ is plotted as well in Fig. 3(c) and reproduces the numerical calculation. $\Gamma$ depends on the ratio between $H_{peak}$ and $H_{th}$ and is only weekly dependent on $H_{peak}$. Namely, when $H_{peak} \gg H_{th}$, the circular trajectory of $\vec{M}$ in the $x - y$ plane persists longer after $t_{peak}$, but as the amplitude of the pulse decays below $H_{th}/\sqrt{2}$, $\vec{M}$ is driven out of the $x - y$ plane and the reversal takes place (see Supplemental Material Note 5). This analysis also holds for LCP pulses, which result in an opposite reversal of $\vec{M}$, as shown in Fig. 3(d).

To summarize, in this work we demonstrated that the control of the magnetization by an optical field arises from first principles by introducing the magnetic part of the optical radiation to the LLG equation. This was seen from the comparison between the case where $\eta \ll 1$ and the case of $\eta = 1$. Using the TLS model, we demonstrated the coupling between the optical helicity state and the polarity of the longitudinal torque. A quantitative analysis of the optically induced torque revealed that it can be comparable to that observed in experiments.

**Figure 1**

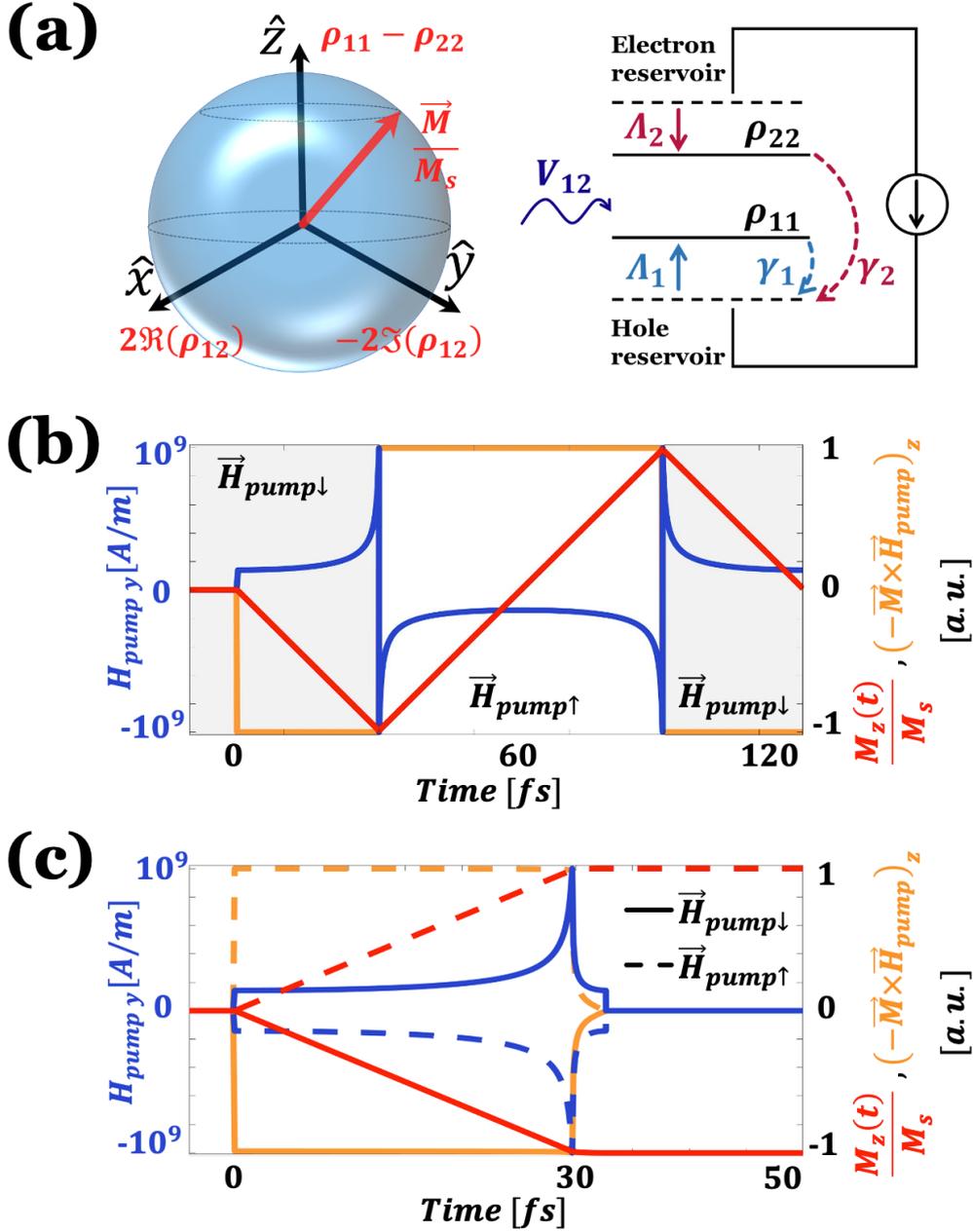

Fig. 1. (a) Left panel: Illustration of $\vec{M}$ on the Bloch sphere. Right panel: Illustration of the electrically pumped TLS. (b) Interaction with $\vec{H}_{pump\downarrow\uparrow}$ of Eq. (5). The Figure illustrates the temporal plots of $M_z/M_s$, $\vec{H}_{pump\downarrow\uparrow,y}$ and $\left(-\vec{M}\times\vec{H}\right)_z$ normalized to unity. (c) Interaction with $\vec{H}_{pump\downarrow\uparrow}$ and a more realistic trailing edge, for the same conditions in (b). Full lines correspond to $\vec{H}_{pump\downarrow}$ and dashed lines correspond to $\vec{H}_{pump\uparrow}$.

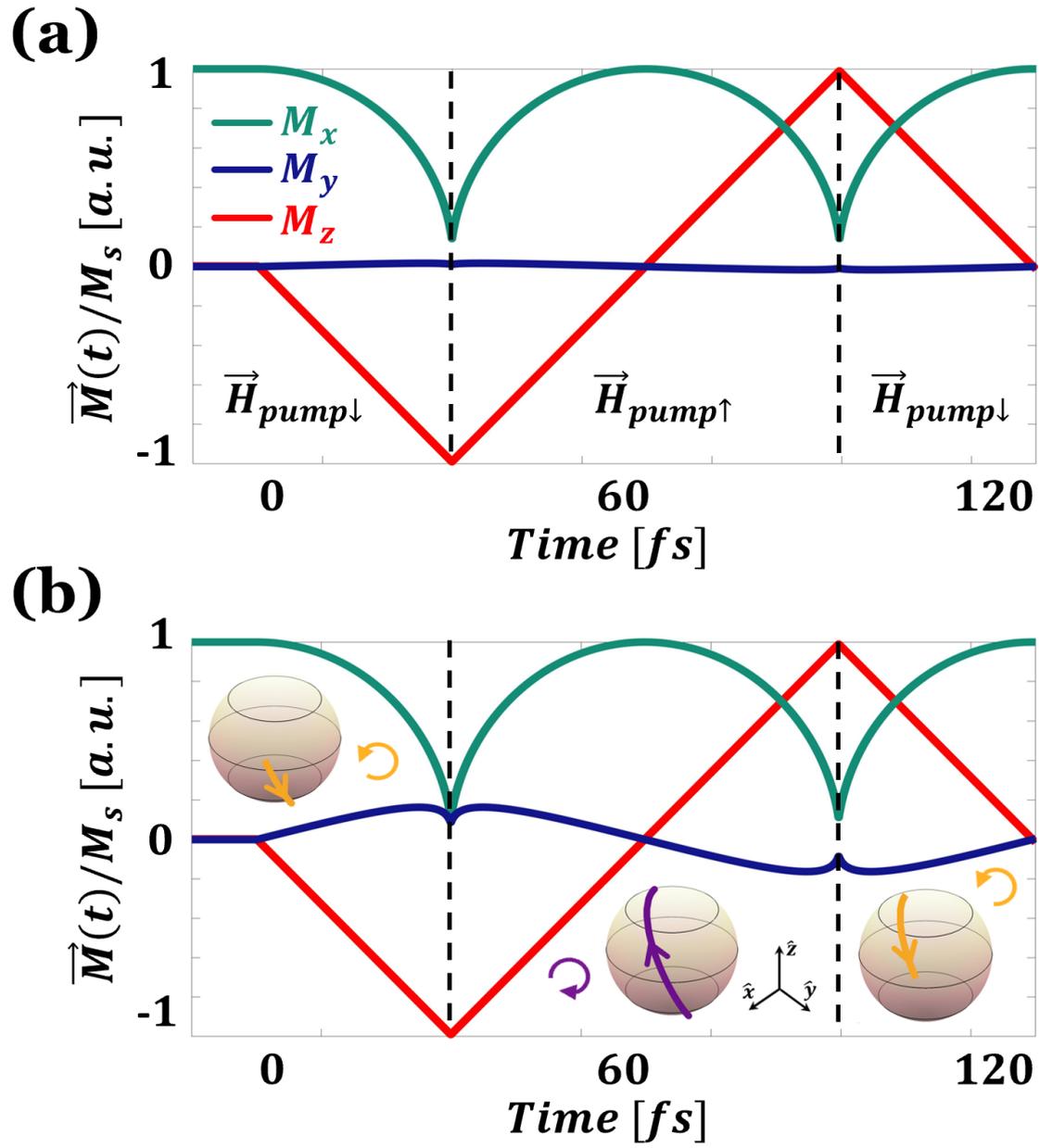

**Fig. 2.** Temporal evolution of the components of $\vec{M}$ under the influence of alternating $\vec{H}_{pump\downarrow}$ and $\vec{H}_{pump\uparrow}$ for (a) small and (b) large damping. Black dashed lines indicate the alternation between $\vec{H}_{pump\downarrow}$ and $\vec{H}_{pump\uparrow}$.

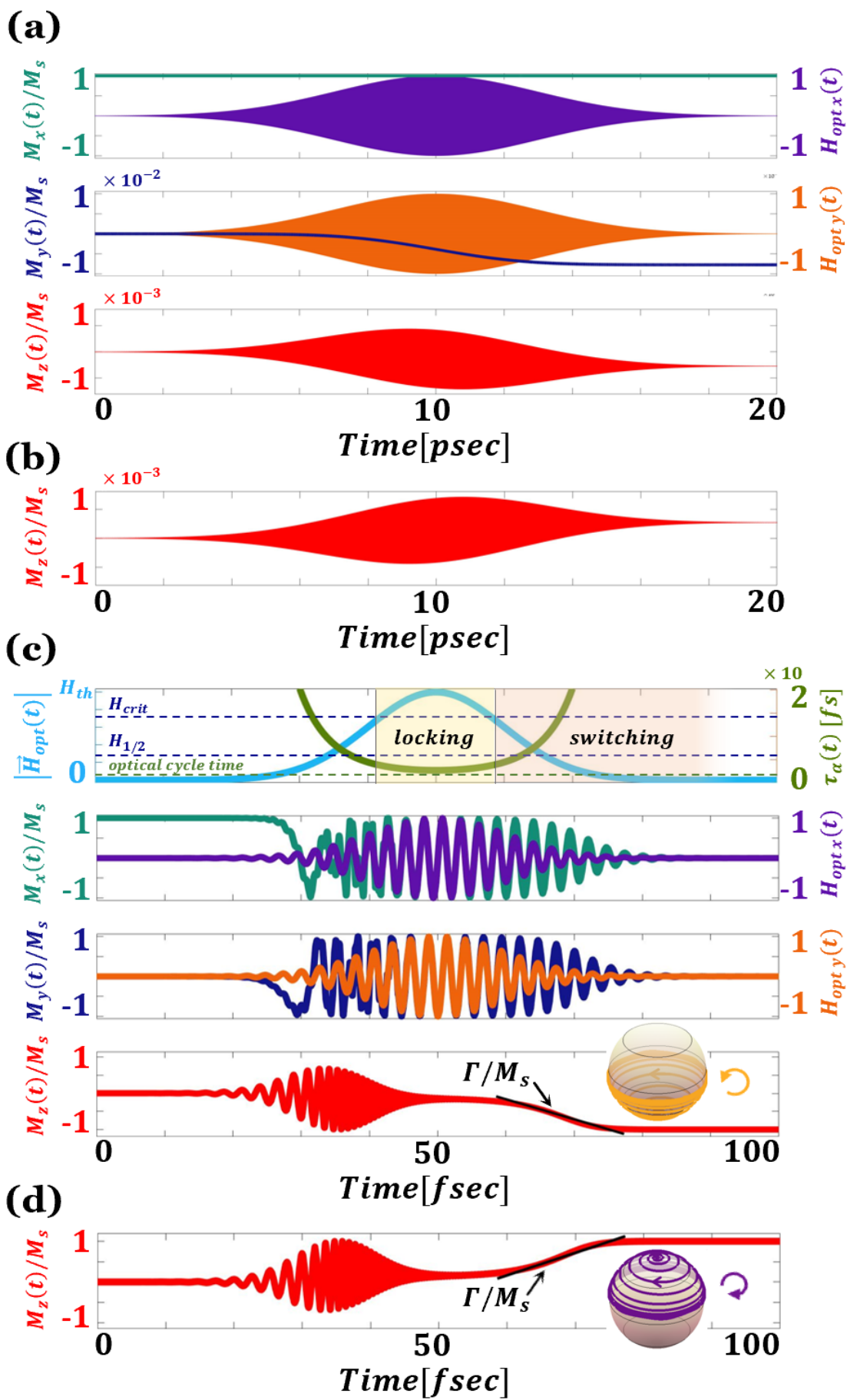

Fig. 3. (a) Magnetization reversal induced by an RCP Gaussian pulse for $\eta = 2.5 \cdot 10^{-4}$. Top and middle panels depict the temporal evolution of the $x$ and $y$ components of $\vec{M}$ and $\vec{H}_{opt}$ in normalized units. Bottom panel depicts $M_z/M_s$. (b) $M_z/M_s$, for the application of an LCP pulse. (c) Magnetization reversal induced by an RCP Gaussian pulse for $\eta = 1$. Top panel presents the temporal behavior of $|\vec{H}_{opt}|$ and $\tau_\alpha$, where $H_{crit} = H_{th}/\sqrt{2}$ and $H_{1/2} = 0.27 H_{th}$. (d) $M_z/M_s$, for the application of an LCP pulse. In (c) and (d), black solid lines represent the analytical solution of $\Gamma/M_s$.